\input harvmac
\sequentialequations



\def\TL{\hfil$\displaystyle{##}$}
\def\TR{$\displaystyle{{}##}$\hfil}
\def\TC{\hfil$\displaystyle{##}$\hfil}
\def\TT{\hbox{##}}



\def\comment#1{}
\def\fixit#1{}

\def\tf#1#2{{\textstyle{#1 \over #2}}}
\def\df#1#2{{\displaystyle{#1 \over #2}}}


\def\sech{\mathop{\rm sech}\nolimits}
\def\Vol{\mathop{\rm Vol}\nolimits}

\def\diag{\mathop{\rm diag}\nolimits}

\def\lsim{\mathrel{\mathstrut\smash{\ooalign{\raise2.5pt\hbox{$<$}\cr\lower2.5pt\hbox{$\sim$}}}}}
\def\gsim{\mathrel{\mathstrut\smash{\ooalign{\raise2.5pt\hbox{$>$}\cr\lower2.5pt\hbox{$\sim$}}}}}


\def\slashed#1{\ooalign{\hfil\hfil/\hfil\cr $#1$}}

\def\sqr#1#2{{\vcenter{\vbox{\hrule height.#2pt
         \hbox{\vrule width.#2pt height#1pt \kern#1pt
            \vrule width.#2pt}
         \hrule height.#2pt}}}}
\def\square{\mathop{\mathchoice\sqr56\sqr56\sqr{3.75}4\sqr34}\nolimits}




\def\eff{{\rm eff}}
\def\abs{{\rm abs}}
\def\hc{{\rm h.c.}}
\def\+{^\dagger}



\def\overleftrightarrow#1{\vbox{\ialign{##\crcr
     $\leftrightarrow$\crcr\noalign{\kern-0pt\nointerlineskip}
     $\hfil\displaystyle{#1}\hfil$\crcr}}}
\def\em{\it}   




\def\cm{[1]}
\def\ms{[2]}
\def\brekEx{[3]}
\def\brekNonEx{[4]}
\def\sv{[5]}
\def\dmw{[6]}
\def\dmOne{[7]}
\def\dmTwo{[8]}
\def\gkOne{[9]}
\def\mast{[10]}
\def\gkgrey{[11]}
\def\cgkt{[12]}
\def\kk{[13]}
\def\ja{[14]}
\def\krt{[15]}
\def\htr{[16]}
\def\dkt{[17]}
\def\kkTwo{[18]}
\def\mathur{[19]}
\def\gunp{[20]}
\def\gukt{[21]}
\def\semicl{2}
\def\ScatteringSol{(1)}
\def\OpticalTheorem{(2)}
\def\Watson{[22]}
\def\NeumannSeries{(3)}
\def\LegendreExpand{(4)}
\def\ChangeFactor{(5)}
\def\PartialWave{(6)}
\def\TotalScatter{(7)}
\def\SigmaScatter{(8)}
\def\SigmaAbsGen{(9)}
\def\SigmaAbsFF{(10)}
\def\MetricNE{(11)}
\def\ThermoQs{(12)}
\def\ExactNE{(13)}
\def\Unruh{[23]}
\def\hmf{[24]}
\def\EtaSols{(14)}
\def\EtaDef{(15)}
\def\EtaMatch{(16)}
\def\CDef{(17)}
\def\ProbAbs{(18)}
\def\SigmaGR{(19)}
\def\finn{[25]}
\def\SigmaNonEx{(20)}
\def\TLR{(21)}
\def\Dbrane{3}
\def\cederOne{[26]}
\def\cederTwo{[27]}
\def\aps{[28]}
\def\atWhich{[29]}
\def\VintSure{(22)}
\def\juan{[30]}
\def\gforth{[31]}
\def\CapTeff{(23)}
\def\tandT{(24)}
\def\VintGuess{(25)}
\def\GammaMs{(26)}
\def\TauMs{(27)}
\def\PsiDs{(28)}
\def\VintGuessTwo{(29)}
\def\SKinetic{(30)}
\def\ModeExpansions{(31)}
\def\dmI{[32]}
\def\ModeCommutators{(32)}
\def\TildeIF{(33)}
\def\OneNZTerm{(34)}
\def\ONZTEnhancedGolden{(35)}
\def\Rhos{(36)}
\def\MoreTildeF{(37)}
\def\NonZeroTerms{(38)}
\def\EnhancedGolden{(39)}
\def\ContApprox{(40)}
\def\SigmaPLR{(41)}
\def\ILeft{(42)}
\def\SechIntegrals{(43)}
\def\ILeftComp{(44)}
\def\IRight{(45)}
\def\SigmaD{(46)}
\def\TLt{(47)}
\def\FixedTension{(48)}
\def\CNorm{(49)}
\def\SigmaDNonEx{(50)}
\def\TLtNonEx{(51)}
\def\llimits{4}
\def\klebThree{[33]}
\def\kvk{[34]}
\def\ThreeBounds{(52)}
\def\NRels{(53)}
\def\juanLow{[35]}
\def\StandardLims{(54)}
\def\FatLim{(55)}
\def\DesiredLims{(56)}
\def\Conclusion{5}



\def\jref{}   
\def\href#1#2{#2}



\lref\hmf{{\it Handbook of Mathematical Functions}, M.~Abramowitz and
I.~A. Stegun, eds. (US Government Printing Office, Washington, DC,
1964).}

\lref\blatt{J.~M. Blatt and V.~F. Weisskopf, {\it Theoretical Nuclear
Physics} (New York: Wiley, 1952), p.~520.}

\lref\bd{P.~C.~W. Birrell and N.~D. Davies, {\it Quantum Fields in Curved
Space} (Cambridge, UK: Cambridge University Press, 1982).}

\lref\carrier{G.~F. Carrier, M.~Krook, and C.~E. Pearson, {\it Functions
of a complex variable} (Ithaca: Hod Books, 1983).}

\lref\Chandra{S.~Chandrasekhar, ``On the equations governing the
perturbations of the Reissner-Nordstrom black hole,'' {\it
Proc. R. Soc. Lond.} A.~365 (1979) 453-465.}

\lref\Fut{J.~F. Futterman, F.~Handler, and R.~Matzner, {\it Scattering
from Black Holes} (Cambridge, UK: Cambridge University Press, 1988).}

\lref\gibbOne{
G.~W. Gibbons, ``Antigravitating black hole solitons with scalar hair
  in $N = 4$ supergravity,'' {\em Nucl. Phys.} {\bf B207} (1982) 337.}

\lref\gibbTwo{P.~Breitenlohner, D.~Maison and G.~Gibbons, {\it
Commun.~Math.~Phys.}~120 (1988) 295.}\jref

\lref\GR{I.~S. Gradshteyn and I.~M. Ryzhik, {\it Table of Integrals,
Series, and Products}, Fifth Edition, A.~Jeffrey, ed. (Academic Press:
San Diego, 1994).}

\lref\gold{J.~N. Goldberg, A.~J. Macfarlane, E.~T. Newman, F.~Rohrlich,
and E.~C.~G. Sudarshan, ``Spin-$s$ Spherical Harmonics and
$\slashed\partial$,'' {\em J.~Math. Phys.} {\bf 8} (1967) 2155-2161.}

\lref\gunp{S.~S. Gubser, unpublished notes, November 1996.}

\lref\gforth{S.~S. Gubser, forthcoming.}

\lref\np{E.~T. Newman and R.~Penrose, ``An approach to gravitational
radiation by a method of spin coefficients,'' {\em J.~Math. Phys.} {\bf 3}
(1962) 566-578; erratum {\bf 4}, 998.}

\lref\ageOne{D.~Page, ``Particle emission rates from a black hole:
massless particles from an uncharged, nonrotating hole,'' {\em
Phys. Rev.} {\bf D13} (1976) 198-206.}

\lref\ageTwo{D.~Page, {\em Phys. Rev.} {\bf D15} (1976) 3260.}\jref

\lref\pr{R.~Penrose and W.~Rindler, {\it Spacetime and Spinors},
Vol.~1 (Cambridge, UK: Cambridge Univerisity Press, 1984).}

\lref\rs{W.~Rarita and J.~Schwinger, ``On a theory of particles with
half-integral spin,'' {\em Phys. Rev.} {\bf 60} (1941) 61.}

\lref\teuk{S.~A. Teukolsky, ``Rotating black holes: separable wave
equations for gravitational and electromagnetic perturbations,''
{\em Phys. Rev. Lett.} {\bf 29} (1972) 1114-1118.}

\lref\teukI{S.~A. Teukolsky, ``Perturbations of a rotating black
hole I'' {\em Astrophys.~J.} {\bf 185} (1973) 635-647.}

\lref\pktIII{P.~K. Townsend, Part~III lectures given at Cambridge
University, Lent 1995.}

\lref\Unruh{W.~G. Unruh, ``Absorption cross section of small black
holes,'' {\em Phys. Rev.} {\bf D14} (1976) 3251-3258.}

\lref\Wald{R.~M. Wald, {\em General Relativity} (Chicago: The
University of Chicago Press, 1984).}

\lref\Watson{G.N.~Watson, {\it Theory of Bessel Functions}, second
edition (Cambridge, UK: Cambridge University Press, 1962).}

\lref\weinbergI{S.~Weinberg, {\it The Quantum Theory of Fields}, Vol.~1
(Cambridge, UK: Cambridge University Press, 1995).}


\lref\brekEx{
J.~C. Breckenridge, R.~C. Myers, A.~W. Peet, and C.~Vafa, ``D-branes and
  spinning black holes,'' {\em Phys. Lett.} {\bf B391} (1996) 93,
  \href{http://xxx.lanl.gov/abs/hep-th/9602065}{{\tt hep-th/9602065}}.}

\lref\brekNonEx{
J.~C. Breckenridge {\em et.~al.}, ``Macroscopic and microscopic entropy of near
  extremal spinning black holes,'' {\em Phys. Lett.} {\bf B381} (1996)
  423--426, \href{http://xxx.lanl.gov/abs/hep-th/9603078}{{\tt
  hep-th/9603078}}.}

\lref\bho{
E.~Bergshoeff, C.~Hull, and T.~Ortin, ``Duality in the type II superstring
  effective action,'' {\em Nucl. Phys.} {\bf B451} (1995) 547--578,
  \href{http://xxx.lanl.gov/abs/hep-th/9504081}{{\tt hep-th/9504081}}.}

\lref\cgkt{
C.~G. Callan, S.~S. Gubser, I.~R. Klebanov, and A.~A. Tseytlin, ``Absorption
  of fixed scalars and the D-brane approach to black holes,'' {\em Nucl. Phys.}
  {\bf B489} (1997) 65--94, \href{http://xxx.lanl.gov/abs/hep-th/9610172}{{\tt
  hep-th/9610172}}.}

\lref\cm{
C.~G. Callan and J.~M. Maldacena, ``D-brane approach to black hole
  quantum mechanics,'' {\em Nucl. Phys.} {\bf B472} (1996) 591--610,
  \href{http://xxx.lanl.gov/abs/hep-th/9602043}{{\tt hep-th/9602043}}.}

\lref\csfOne{
E.~Cremmer, J.~Scherk, and S.~Ferrara, {\em Phys. Lett.} {\bf B68}
 (1977) 234.}\jref

\lref\csfTwo{
E.~Cremmer, J.~Scherk, and S.~Ferrara, ``SU(4) invariant supergravity
 theory,'' {\em Phys. Lett.} {\bf B74} (1978) 61--64.}

\lref\cs{
E.~Cremmer and J.~Scherk, ``Algebraic simplifications in supergravity 
 theories,'' {\em Nucl. Phys.} {\bf B127} (1977) 259--268.}

\lref\CY{
M.~Cvetic and D.~Youm, ``BPS saturated and nonextreme states in abelian
  Kaluza-Klein theory and effective N=4 supersymmetric string vacua,''
  \href{http://xxx.lanl.gov/abs/hep-th/9508058}{{\tt hep-th/9508058}}.}

\lref\CYY{
M.~Cvetic and D.~Youm, ``General rotating five-dimensional black holes of
  toroidally compactified heterotic string,'' {\em Nucl. Phys.} {\bf B476}
  (1996) 118--132, \href{http://xxx.lanl.gov/abs/hep-th/9603100}{{\tt
  hep-th/9603100}}.}

\lref\CT{
M.~Cvetic and A.~A. Tseytlin, ``Nonextreme black holes from nonextreme
  intersecting M-branes,'' {\em Nucl. Phys.} {\bf B478} (1996) 181--198,
  \href{http://xxx.lanl.gov/abs/hep-th/9606033}{{\tt hep-th/9606033}}.}

\lref\CTT{
M.~Cvetic and A.~A. Tseytlin, ``Solitonic strings and BPS saturated dyonic
  black holes,'' {\em Phys. Rev.} {\bf D53} (1996) 5619--5633,
  \href{http://xxx.lanl.gov/abs/hep-th/9512031}{{\tt hep-th/9512031}}.}

\lref\adas{
A.~Das, ``SO(4) invariant extended supergravity,'' {\em Phys. Rev.}
 {\bf D15} (1977) 2805.}

\lref\dmOne{
S.~R. Das and S.~D. Mathur, ``Comparing decay rates for black holes and
  D-branes,'' {\em Nucl. Phys.} {\bf B478} (1996) 561--576,
  \href{http://xxx.lanl.gov/abs/hep-th/9606185}{{\tt hep-th/9606185}}.}

\lref\dmTwo{
S.~R. Das and S.~D. Mathur, ``Interactions involving D-branes,'' {\em Nucl.
  Phys.} {\bf B482} (1996) 153--172,
  \href{http://xxx.lanl.gov/abs/hep-th/9607149}{{\tt hep-th/9607149}}.}

\lref\dgm{
S.~R. Das, G.~Gibbons, and S.~D. Mathur, ``Universality of low-energy
  absorption cross-sections for black holes,'' {\em Phys. Rev. Lett.} {\bf 78}
  (1997) 417--419, \href{http://xxx.lanl.gov/abs/hep-th/9609052}{{\tt
  hep-th/9609052}}.}

\lref\dmI{
S.~R. Das and S.~D. Mathur, ``Excitations of D strings, entropy and duality,''
  {\em Phys. Lett.} {\bf B375} (1996) 103--110,
  \href{http://xxx.lanl.gov/abs/hep-th/9601152}{{\tt hep-th/9601152}}.}

\lref\dmw{
A.~Dhar, G.~Mandal, and S.~R. Wadia, ``Absorption vs. decay of black holes in
  string theory and T symmetry,'' {\em Phys. Lett.} {\bf B388} (1996) 51--59,
  \href{http://xxx.lanl.gov/abs/hep-th/9605234}{{\tt hep-th/9605234}}.}

\lref\fkOne{
S.~Ferrara and R.~Kallosh, ``Supersymmetry and attractors,'' {\em Phys. Rev.}
  {\bf D54} (1996) 1514--1524,
  \href{http://xxx.lanl.gov/abs/hep-th/9602136}{{\tt hep-th/9602136}}.}

\lref\fkTwo{
S.~Ferrara and R.~Kallosh, ``Universality of supersymmetric attractors,'' {\em
  Phys. Rev.} {\bf D54} (1996) 1525--1534,
  \href{http://xxx.lanl.gov/abs/hep-th/9603090}{{\tt hep-th/9603090}}.}

\lref\fks{
S.~Ferrara, R.~Kallosh, and A.~Strominger, ``N=2 extremal black holes,'' {\em
  Phys. Rev.} {\bf D52} (1995) 5412--5416,
  \href{http://xxx.lanl.gov/abs/hep-th/9508072}{{\tt hep-th/9508072}}.}

\lref\gkt{
J.~P. Gauntlett, D.~A. Kastor, and J.~Traschen, ``Overlapping branes in M
  theory,'' {\em Nucl. Phys.} {\bf B478} (1996) 544--560,
  \href{http://xxx.lanl.gov/abs/hep-th/9604179}{{\tt hep-th/9604179}}.}

\lref\gkk{
G.~Gibbons, R.~Kallosh, and B.~Kol, ``Moduli, scalar charges, and the first law
  of black hole thermodynamics,'' {\em Phys. Rev. Lett.} {\bf 77} (1996)
  4992--4995, \href{http://xxx.lanl.gov/abs/hep-th/9607108}{{\tt
  hep-th/9607108}}.}

\lref\gso{
F.~Gliozzi, J.~Scherk, and D.~Olive, ``Supersymmetry, supergravity
 theories and the dual spinor model,'' {\em Nucl. Phys.} {\bf B122}
 (1977) 253-290.}

\lref\gil{
G.~Gilbert, ``On the perturbations of string theoretic black holes,''
  \href{http://xxx.lanl.gov/abs/hep-th/9108012}{{\tt hep-th/9108012}}.}

\lref\gregOne{
R.~Gregory and R.~Laflamme, ``Evidence for stability of extremal black
  p-branes,'' {\em Phys. Rev.} {\bf D51} (1995) 305--309,
  \href{http://xxx.lanl.gov/abs/hep-th/9410050}{{\tt hep-th/9410050}}.}

\lref\gregTwo{
R.~Gregory and R.~Laflamme, ``The instability of charged black strings and
  p-branes,'' {\em Nucl. Phys.} {\bf B428} (1994) 399--434,
  \href{http://xxx.lanl.gov/abs/hep-th/9404071}{{\tt hep-th/9404071}}.}

\lref\hw{
C.~F.~E. Holzhey and F.~Wilczek, ``Black holes as elementary particles,'' {\em
  Nucl. Phys.} {\bf B380} (1992) 447--477,
  \href{http://xxx.lanl.gov/abs/hep-th/9202014}{{\tt hep-th/9202014}}.}

\lref\gkp{
S.~S. Gubser, I.~R. Klebanov, and A.~W. Peet, ``Entropy and temperature of
  black 3-branes,'' {\em Phys. Rev.} {\bf D54} (1996) 3915--3919,
  \href{http://xxx.lanl.gov/abs/hep-th/9602135}{{\tt hep-th/9602135}}.}

\lref\gkOne{
S.~S. Gubser and I.~R. Klebanov, ``Emission of charged particles from
  four-dimensional and five-dimensional black holes,'' {\em Nucl. Phys.} {\bf
  B482} (1996) 173--186, \href{http://xxx.lanl.gov/abs/hep-th/9608108}{{\tt
  hep-th/9608108}}.}

\lref\gkgrey{
S.~S. Gubser and I.~R. Klebanov, ``Four-dimensional grey body factors and the
  effective string,'' {\em Phys. Rev. Lett.} {\bf 77} (1996) 4491--4494,
  \href{http://xxx.lanl.gov/abs/hep-th/9609076}{{\tt hep-th/9609076}}.}

\lref\hkrs{
E.~Halyo, B.~Kol, A.~Rajaraman, and L.~Susskind, ``Counting schwarzchild and
  charged black holes,'' \href{http://xxx.lanl.gov/abs/hep-th/9609075}{{\tt
  hep-th/9609075}}.}

\lref\halyo{
E.~Halyo, ``Reissner-Nordstrom black holes and strings with rescaled tension,''
  \href{http://xxx.lanl.gov/abs/hep-th/9610068}{{\tt hep-th/9610068}}.}

\lref\hms{
G.~T. Horowitz, J.~M. Maldacena, and A.~Strominger, ``Nonextremal black hole
  microstates and U duality,'' {\em Phys. Lett.} {\bf B383} (1996) 151--159,
  \href{http://xxx.lanl.gov/abs/hep-th/9603109}{{\tt hep-th/9603109}}.}

\lref\HS{
G.~T. Horowitz and A.~Strominger, ``Counting states of near extremal black
  holes,'' {\em Phys. Rev. Lett.} {\bf 77} (1996) 2368--2371,
  \href{http://xxx.lanl.gov/abs/hep-th/9602051}{{\tt hep-th/9602051}}.}

\lref\hp{
G.~T. Horowitz and J.~Polchinski, ``A correspondence principle for black holes
  and strings,'' \href{http://xxx.lanl.gov/abs/hep-th/9612146}{{\tt
  hep-th/9612146}}.}

\lref\myers{
C.~V. Johnson, R.~R. Khuri, and R.~C. Myers, ``Entropy of 4-d extremal black
  holes,'' {\em Phys. Lett.} {\bf B378} (1996) 78--86,
  \href{http://xxx.lanl.gov/abs/hep-th/9603061}{{\tt hep-th/9603061}}.}

\lref\klopp{
R.~Kallosh, A.~Linde, T.~Ortin, A.~Peet, and A.~V. Proeyen, ``Supersymmetry as
  a cosmic censor,'' {\em Phys. Rev.} {\bf D46} (1992) 5278--5302,
  \href{http://xxx.lanl.gov/abs/hep-th/9205027}{{\tt hep-th/9205027}}.}

\lref\khuri{
R.~R. Khuri, ``Manton scattering of string solitons,'' {\em Nucl. Phys.} {\bf
  B376} (1992) 350--364.}

\lref\ktI{
I.~R. Klebanov and A.~A. Tseytlin, ``Near extremal black hole entropy and
  fluctuating three-branes,'' {\em Nucl. Phys.} {\bf B479} (1996) 319--335,
  \href{http://xxx.lanl.gov/abs/hep-th/9607107}{{\tt hep-th/9607107}}.}

\lref\KT{
I.~R. Klebanov and A.~A. Tseytlin, ``Intersecting M-branes as four-dimensional
  black holes,'' {\em Nucl. Phys.} {\bf B475} (1996) 179--192,
  \href{http://xxx.lanl.gov/abs/hep-th/9604166}{{\tt hep-th/9604166}}.}

\lref\kr{
B.~Kol and A.~Rajaraman, ``Fixed scalars and suppression of Hawking
  evaporation,'' \href{http://xxx.lanl.gov/abs/hep-th/9608126}{{\tt
  hep-th/9608126}}.}

\lref\kleb{
I.~R. Klebanov, ``World volume approach to absorption by nondilatonic branes,''
  \href{http://xxx.lanl.gov/abs/hep-th/9702076}{{\tt hep-th/9702076}}.}

\lref\kk{
I.~R. Klebanov and M.~Krasnitz, ``Fixed scalar gray body factors in
  five-dimensions and four-dimensions,'' {\em Phys. Rev.} {\bf D55} (1997)
  3250--3254, \href{http://xxx.lanl.gov/abs/hep-th/9612051}{{\tt
  hep-th/9612051}}.}

\lref\kkTwo{
M.~Krasnitz and I.~R. Klebanov, ``Testing effective string models of black
  holes with fixed scalars,''
  \href{http://xxx.lanl.gov/abs/hep-th/9703216}{{\tt hep-th/9703216}}.}

\lref\finn{
F.~Larsen, ``A string model of black hole microstates,''
  \href{http://xxx.lanl.gov/abs/hep-th/9702153}{{\tt hep-th/9702153}}.}\jref

\lref\lwOne{
F.~Larsen and F.~Wilczek, ``Internal structure of black holes,'' {\em Phys.
  Lett.} {\bf B375} (1996) 37--42,
  \href{http://xxx.lanl.gov/abs/hep-th/9511064}{{\tt hep-th/9511064}}.}

\lref\lwTwo{
F.~Larsen and F.~Wilczek, ``Classical hair in string theory. 2: Explicit
  calculations,'' {\em Nucl. Phys.} {\bf B488} (1997) 261--281,
  \href{http://xxx.lanl.gov/abs/hep-th/9609084}{{\tt hep-th/9609084}}.}

\lref\lu{
J.~X. Lu, ``ADM masses for black strings and p-branes,'' {\em Phys. Lett.} {\bf
  B313} (1993) 29--34, \href{http://xxx.lanl.gov/abs/hep-th/9304159}{{\tt
  hep-th/9304159}}.}

\lref\maha{
J.~Maharana and J.~H. Schwarz, ``Noncompact symmetries in string theory,'' {\em
  Nucl. Phys.} {\bf B390} (1993) 3--32,
  \href{http://xxx.lanl.gov/abs/hep-th/9207016}{{\tt hep-th/9207016}}.}

\lref\mast{
J.~Maldacena and A.~Strominger, ``Black hole grey body factors and D-brane
  spectroscopy,'' {\em Phys. Rev.} {\bf D55} (1997) 861--870,
  \href{http://xxx.lanl.gov/abs/hep-th/9609026}{{\tt hep-th/9609026}}.}

\lref\mst{
J.~M. Maldacena and A.~Strominger, ``Statistical entropy of four-dimensional
  extremal black holes,'' {\em Phys. Rev. Lett.} {\bf 77} (1996) 428--429,
  \href{http://xxx.lanl.gov/abs/hep-th/9603060}{{\tt hep-th/9603060}}.}

\lref\ja{
J.~Maldacena and A.~Strominger, ``Universal low-energy dynamics for rotating
  black holes,'' \href{http://xxx.lanl.gov/abs/hep-th/9702015}{{\tt
  hep-th/9702015}}.}

\lref\ms{
J.~M. Maldacena and L.~Susskind, ``D-branes and fat black holes,'' {\em Nucl.
  Phys.} {\bf B475} (1996) 679--690,
  \href{http://xxx.lanl.gov/abs/hep-th/9604042}{{\tt hep-th/9604042}}.}

\lref\juan{
J.~M. Maldacena, ``Statistical entropy of near extremal five-branes,'' {\em
  Nucl. Phys.} {\bf B477} (1996) 168--174,
  \href{http://xxx.lanl.gov/abs/hep-th/9605016}{{\tt hep-th/9605016}}.}

\lref\juanLow{
J.~Maldacena, ``D-branes and near extremal black holes at low-energies,''
  \href{http://xxx.lanl.gov/abs/hep-th/9611125}{{\tt hep-th/9611125}}.}

\lref\mathur{
S.~D. Mathur, ``Absorption of angular momentum by black holes and
 D-branes,'' {\tt hep-th/9704156}.}

\lref\jpTASI{
J.~Polchinski, ``TASI lectures on D-branes,''
  \href{http://xxx.lanl.gov/abs/hep-th/9611050}{{\tt hep-th/9611050}}.}

\lref\schw{
J.~H. Schwarz, ``Covariant field equations of chiral N=2 d = 10 supergravity,''
  {\em Nucl. Phys.} {\bf B226} (1983) 269.}

\lref\sv{
A.~Strominger and C.~Vafa, ``Microscopic origin of the Bekenstein-Hawking
  entropy,'' {\em Phys. Lett.} {\bf B379} (1996) 99--104,
  \href{http://xxx.lanl.gov/abs/hep-th/9601029}{{\tt hep-th/9601029}}.}

\lref\at{
A.~A. Tseytlin, ``Harmonic superpositions of M-branes,'' {\em Nucl. Phys.} {\bf
  B475} (1996) 149--163, \href{http://xxx.lanl.gov/abs/hep-th/9604035}{{\tt
  hep-th/9604035}}.}

\lref\ATT{
A.~A. Tseytlin, ``Extreme dyonic black holes in string theory,'' {\em Mod.
  Phys. Lett.} {\bf A11} (1996) 689--714,
  \href{http://xxx.lanl.gov/abs/hep-th/9601177}{{\tt hep-th/9601177}}.}


\lref\dkt{
F.~Dowker, D.~Kastor, and J.~Traschen, ``U duality, D-branes and black hole
  emission rates: Agreements and disagreements,''
  \href{http://xxx.lanl.gov/abs/hep-th/9702109}{{\tt hep-th/9702109}}.}\jref

\lref\kvk{
E.~Keski-Vakkuri and P.~Kraus, ``Microcanonical D-branes and back reaction,''
  \href{http://xxx.lanl.gov/abs/hep-th/9610045}{{\tt hep-th/9610045}}.}\jref

\lref\km{
I.~R. Klebanov and S.~D. Mathur, ``Black hole grey body factors and absorption
  of scalars by effective strings,''
  \href{http://xxx.lanl.gov/abs/hep-th/9701187}{{\tt hep-th/9701187}}.}\jref

\lref\htr{
S.~W. Hawking and M.~M. Taylor-Robinson, ``Evolution of near extremal black
  holes,'' 
  \href{http://xxx.lanl.gov/abs/hep-th/9702045}{{\tt hep-th/9702045}}.}\jref

\lref\gukt{
S.~S. Gubser, I.~R. Klebanov, and A.~A. Tseytlin, ``String theory and classical
  absorption by three-branes,''
  \href{http://xxx.lanl.gov/abs/hep-th/9703040}{{\tt hep-th/9703040}}.}\jref

\lref\cederOne{
M.~Cederwall, A.~von Gussich, B.~E.~W. Nilsson, and A.~Westerberg, ``The
  Dirichlet super three brane in ten-dimensional type IIB supergravity,'' {\em
  Nucl. Phys.} {\bf B490} (1997) 163,
  \href{http://xxx.lanl.gov/abs/hep-th/9610148}{{\tt hep-th/9610148}}.}

\lref\cederTwo{
M.~Cederwall, A.~von Gussich, B.~E.~W. Nilsson, P.~Sundell, and A.~Westerberg,
  ``The Dirichlet super p-branes in ten-dimensional type IIA and IIB
  supergravity,'' {\em Nucl. Phys.} {\bf B490} (1997) 179--201,
  \href{http://xxx.lanl.gov/abs/hep-th/9611159}{{\tt hep-th/9611159}}.}

\lref\aps{
M.~Aganagic, C.~Popescu, and J.~H. Schwarz, ``D-brane actions with local kappa
  symmetry,'' {\em Phys. Lett.} {\bf B393} (1997) 311--315,
  \href{http://xxx.lanl.gov/abs/hep-th/9610249}{{\tt hep-th/9610249}}.}

\lref\atWhich{
A.~A. Tseytlin, ``On nonabelian generalization of Born-Infeld action in string
  theory,'' \href{http://xxx.lanl.gov/abs/hep-th/9701125}{{\tt
  hep-th/9701125}}.}\jref

\lref\klebThree{
I.~R. Klebanov, ``World volume approach to absorption by nondilatonic branes,''
  \href{http://xxx.lanl.gov/abs/hep-th/9702076}{{\tt hep-th/9702076}}.}

\lref\krt{
I.~R. Klebanov, A.~Rajaraman, and A.~A. Tseytlin, ``Intermediate scalars and
  the effective string model of black holes,''
  \href{http://xxx.lanl.gov/abs/hep-th/9704112}{{\tt hep-th/9704112}}.}

\lref\hlm{
G.~T. Horowitz, D.~A. Lowe, and J.~M. Maldacena, ``Statistical entropy of
  nonextremal four-dimensional black holes and U duality,'' {\em Phys. Rev.
  Lett.} {\bf 77} (1996) 430--433,
  \href{http://xxx.lanl.gov/abs/hep-th/9603195}{{\tt hep-th/9603195}}.}

\lref\cvetic{
M.~Cvetic, ``Properties of black holes in toroidally compactified string
  theory,'' \href{http://xxx.lanl.gov/abs/hep-th/9701152}{{\tt
  hep-th/9701152}}.}

\lref\susskind{
L.~Susskind, ``Some speculations about black hole entropy in string theory,''
  \href{http://xxx.lanl.gov/abs/hep-th/9309145}{{\tt hep-th/9309145}}.}

\lref\asen{
A.~Sen, ``Extremal black holes and elementary string states,'' {\em Mod. Phys.
  Lett.} {\bf A10} (1995) 2081--2094,
  \href{http://xxx.lanl.gov/abs/hep-th/9504147}{{\tt hep-th/9504147}}.}

\lref\juanFour{
J.~M. Maldacena, ``N=2 extremal black holes and intersecting branes,''
  \href{http://xxx.lanl.gov/abs/hep-th/9611163}{{\tt hep-th\slash9611163}}.}

\lref\gpartial{
S.~S. Gubser, I.~R. Klebanov, and A.~A. Tseytlin, ``String theory and classical
  absorption by three-branes,''
  \href{http://xxx.lanl.gov/abs/hep-th/9703040}{{\tt hep-th/9703040}}.}


\Title{
 \vbox{\baselineskip10pt
  \hbox{PUPT-1697}
  \hbox{hep-th/9704195}
 }
}
{
 \vbox{
  \centerline{Can the effective string}
  \vskip 0.1 truein
  \centerline{see higher partial waves?}
 }
}
\vskip -25 true pt

\centerline{
 Steven S.~Gubser,\footnote{$^1$}{e-mail:  ssgubser@viper.princeton.edu}}
\centerline{\it Joseph Henry Laboratories, 
Princeton University, Princeton, NJ  08544}

\centerline {\bf Abstract}
\smallskip
\baselineskip12pt
\noindent

The semi-classical cross-sections for arbitrary partial waves of
ordinary scalars to fall into certain five-dimensional black holes
have a form that seems capable of explanation in terms of the
effective string model.  The kinematics of these processes is analyzed
in detail on the effective string and is shown to reproduce the
correct functional form of the semi-classical cross-sections.  But it
is necessary to choose a peculiar value of the effective string
tension to obtain the correct scaling properties.  Furthermore, the
assumptions of locality and statistics combine to forbid the effective
string from absorbing more than a finite number of partial waves.  The
relation of this limitation to cosmic censorship is discussed.

\Date{April 1997}

\noblackbox
\baselineskip 14pt plus 1pt minus 1pt


\newsec{Introduction}

The D1-brane D5-brane bound state toroidally compactified down to five
dimensions has proven to be one of the most fruitful string theoretic
models of black holes.  Since the original paper of \cm, which
proposed the model as a way to study black hole dynamics in a
manifestly unitary string theoretic framework, and the subsequent work
in \ms\ clarifying the means by which a single multiply wound
effective string arises in a description of the low-energy dynamics,
there have been many exciting papers relating properties of
five-dimensional and four-dimensional black holes to the effective
string.  An explanation of the near-extremal entropy was given in
\refs{\cm,\brekEx,\brekNonEx}, following the ideas originally laid out
in \sv.  Absorption cross-sections and the corresponding Hawking
emission rates were worked out in
\refs{\dmw,\dmOne,\dmTwo,\gkOne,\mast,\gkgrey,\cgkt,\kk,\ja,\krt} yielding
impressive agreement at low energies with the effective string model.
Suggestions that the effective string model may have some flaws or
limitations have arisen in the work of \refs{\htr,\dkt,\kkTwo}.

In a recent paper by Strominger and Maldacena \ja\ it was found from a
general analysis of thermal two-point functions that the effective
string seems capable of explaining the semi-classical absorption
cross-sections for arbitrary partial waves of ordinary scalars.
Subsequent work by Mathur \mathur\ exhibited more detailed agreement
between the effective string model and General Relativity for these
processes.  Ordinary scalars are scalars whose equation of motion in
five-dimensions is $\square \phi = 0$.  The canonical examples of
ordinary scalars are the off-diagonal gravitons $h_{ij}$ with both $i$
and $j$ lying within the D5-brane but perpendicular to the D1-brane.
These are the scalars for whose $s$-wave cross-section full agreement
between General Relativity and the effective string was first achieved
in \dmOne.

The results of \refs{\ja,\mathur} overlap substantially with
unpublished work by myself \gunp.  The present paper is based upon
that work, which extends the results of \mathur\ in certain technical
aspects.  The organization of the paper is as follows.
Section~\semicl\ covers the semi-classical analysis of partial wave
absorption and includes a derivation of a form of the Optical Theorem
for the absorption of scalar particles which was quoted without proof
in \gukt.  Section~\Dbrane\ presents the effective string description
of the same processes, exhibiting along the way a simple method for
performing all the phase space integrals encountered in \mathur.  In
section~\llimits, the limitation on the number of partial waves the
effective string can couple to arising from statistics and locality is
compared with the limitation imposed semi-classically by cosmic
censorship.  Section~\Conclusion\ summarizes the results and indicates
directions for further work.

\newsec{The semi-classical computation}
\seclab\semicl

The quantity that can be conveniently computed using the matching
technique is the absorption probability.  To convert this to an
absorption cross-section, it is necessary to use properties of the
partial wave expansion in four spatial dimensions and to invoke the
Optical Theorem.  The details of this connection were worked out
independently in \mathur, but because the derivation given
below applies for arbitrary dimensions, it seems worthwhile to present
it in full.  In all of what follows, $n = d-1$ will denote the number
of spatial dimensions.

The Optical Theorem for scattering of a scalar field off a spherically
symmetric potential states that if the scattering wave-function has
for large $r$ the asymptotic form
  \eqn\ScatteringSol{
   \phi(\vec{r}) \sim e^{i k x} + f(\theta) {e^{i k r} \over r^{(n-1)/2}}
  }
 (here $x = r \cos \theta$), then the total
cross-section is
  \eqn\OpticalTheorem{
   \sigma_{\rm total} = 
     -2 \left( {2 \pi \over k} \right)^{n-1 \over 2}
      \Re \left( i^{n-1 \over 2} f(0) \right) \ .
  }
 To find the partial wave expansion of \ScatteringSol, it is first
necessary to make a Neumann expansion of the exponential function,
which can be done using Gegenbauer polynomials \refs{\Watson}:
  \eqn\NeumannSeries{\eqalign{
   e^{i r \cos \theta} &= 2^{n/2-1} \Gamma(n/2-1)
    \sum_{\ell = 0}^\infty 
     i^\ell P_\ell(\cos \theta) (\ell + n/2 - 1) 
     {J_{\ell + n/2 - 1}(r) \over r^{n/2 - 1}} \cr
   P_\ell(\cos \theta) &= \sum_{m=0}^{\lfloor \ell/2 \rfloor} 
    (-1)^m 2^{\ell - 2m} 
    {\Gamma(\ell + n/2 - 1 - m) \over 
     \Gamma(n/2 - 1) m! (\ell - 2m)!} 
    \cos^{\ell - 2m} \theta
  }}
 The $P_\ell(\cos\theta)$ are just the Legendre polynomials when
$n=3$.  For arbitrary $n$, they can be defined by the expansion
  \eqn\LegendreExpand{
   (1 - 2 a \cos \theta + a^2)^{1-n/2} = 
    \sum_{\ell=0}^\infty P_\ell(\cos \theta) a^\ell \ .
  }
 An alternate normalization proves more convenient:
  \eqn\ChangeFactor{
   \tilde{P}_\ell(\cos\theta) = \sqrt{2 \over \pi} 2^{n/2-1} 
    \Gamma(n/2-1) (\ell + n/2 - 1) P_\ell(\cos \theta) \ .
  }
 Using asymptotic properties of Bessel functions, one can now write
down the partial wave expansion of \ScatteringSol\ as
  \eqn\PartialWave{
   e^{i k x} + f(\theta) {e^{i k r} \over r^{(n-1)/2}} 
    \sim \sum_{\ell = 0}^\infty \tf{1}{2} \tilde{P}_\ell(\cos \theta)
     {S_\ell e^{i k r} + (-1)^\ell i^{n-1} e^{-i k r} \over
      (i k r)^{(n-1)/2}} \ .
  }
 When $f(\theta) = 0$ identically, $S_\ell = 1$ for all $\ell$.

The absorption cross-section for the $\ell^{\rm th}$ partial wave can
now be computed as the difference between the total $\ell$-wave
cross-section computed via the Optical Theorem,
  \eqn\TotalScatter{
   \sigma^\ell_{\rm total} = 
    -\left( {\sqrt{2 \pi} \over k} \right)^{n-1} 
     \tilde{P}_\ell(1) \Re (S_\ell - 1) \ ,
  }
 and the $\ell$-wave scattering cross-section,
  \eqn\SigmaScatter{
   \sigma^\ell_{\rm scattered} = (\Vol S^{n-2}) 
    {|S_\ell - 1|^2 \over 4 k^{n-1}} 
    \int_0^\pi d\theta \, \sin^{n-2} \theta \,
     \tilde{P}_\ell(\cos \theta)^2 \ .
  }
 The final result,
  \eqn\SigmaAbsGen{
   \sigma^\ell_{\rm abs} = {2^{n-2} \pi^{n/2-1} \over k^{n-1}}
    \Gamma(n/2-1) (\ell + n/2 - 1) {\ell+n-3 \choose \ell} 
    \left( 1 - |S_\ell|^2 \right) \ ,
  }
 relates the absorption cross-section $\sigma^\ell_{\rm abs}$ to the
absorption probability $1 - |S_\ell|^2$.  The results for $n=3$ and
$n=4$ are
  \eqn\SigmaAbsFF{\vcenter{\openup1\jot
   \halign{\strut\span\TL & \span\TR & \span\TT \cr 
    \sigma^\ell_{\rm abs} &= {\pi \over k^2} (2 \ell + 1) 
     \left( 1 - |S_\ell|^2 \right) & 
     \qquad in three spatial dimensions \cr
    \sigma^\ell_{\rm abs} &= {4 \pi \over k^3} (\ell + 1)^2
     \left( 1 - |S_\ell|^2 \right) &
     \qquad in four spatial dimensions. \cr
  }}}

With these results in hand, let us proceed to the semi-classical
computation of the cross-section for an ordinary scalar $\phi$ in the
$\ell^{\rm th}$ partial wave to be absorbed into a black hole.  It is
hoped that this piece of ``spectroscopic data'' will be illuminating
of the form of the effective string action.

An ordinary scalar is one whose equation of motion is just the Laplace
equation following from the black hole metric, which in five
dimensions is
  \eqn\MetricNE{\eqalign{
   ds^2 &= -F^{-2/3} h dt^2 + 
    F^{1/3} \left( h^{-1} dr^2 + r^2 d \Omega_{S^3}^2 \right) \cr
   F &= f_1 f_5 f_K = \left( 1 + {r_1^2 \over r^2} \right)
                      \left( 1 + {r_5^2 \over r^2} \right)
                      \left( 1 + {r_K^2 \over r^2} \right) \cr
   h &= 1 - {r_0^2 \over r^2} \ .
  }}
 The mass, entropy, Hawking temperature, $U(1)$ charges, and
characteristic radii are conveniently parameterized as
  \eqn\ThermoQs{\vcenter{\openup1\jot
    \halign{\strut\span\TC\cr
     M = {\pi \over 8} r_0^2 \sum_{i=1,5,K} \cosh 2 \sigma_i \qquad
     S = {\pi^2 \over 2} r_0^3 \prod_{i=1,5,K} \cosh \sigma_i \qquad
     \beta_H = 2 \pi r_0 \prod_{i=1,5,K} \cosh \sigma_i  \cr 
     Q_i = {r_0^2 \over 2} \sinh 2\sigma_i  \qquad
     r_i = r_0 \sinh \sigma_i  \cr
  }}}
 in five-dimensional Planck units.  Using a separation of variables 
  $\phi = e^{-i \omega t} P_\ell(\cos \theta) R(r)$, 
 one can extract from the Laplace equation $\square \phi = 0$ the
radial equation
  \eqn\ExactNE{
   \left[ (h r^3 \partial_r)^2 + r^6 F \omega^2 - 
    r^4 h \ell (\ell + 2) \right] R = 0 \ .
  }
 Because of the left-right symmetry of the effective string
description for five-dimensional black holes, the absorption
cross-section for the near-extremal case provides essentially no more
information about the effective string than the extremal case does.
In the interest of a simple presentation, I will therefore restrict my
calculations in both this section and the next to the extremal case.
The near-extremal generalizations of the results are summarized at the
end of each section.

In the near horizon region (denoted ${\bf I}$ for consistency with the
literature \refs{\Unruh,\gkgrey,\cgkt}), \ExactNE\ for an extremal
black hole can be approximated by a Coulomb equation \refs{\hmf} in
the variable
  $y = (r_1 r_5 r_K \omega) / (2 r^2)$,
 while in the far horizon region ${\bf III}$ it can be approximated as
a Bessel equation:
  \eqn\EtaSols{\vcenter{\openup1\jot
   \halign{\strut\span\TL & \span\TL & \span\TR & \span\TT & 
                 \span\TL & \span\TR\cr
   {\bf I.}\ \ &\left[ \partial_y^2 + 1 - 
      \df{2\eta}{y} - \df{\ell (\ell + 2)/4}{y^2} \right] R_{\bf I} &= 0 &
    \qquad\ & R_{\bf I} &= G_{\ell/2}(y) + i F_{\ell/2}(y) \cr
   {\bf III.} \ \ &\left[ (r^3 \partial_r)^2 + r^6 \omega^2 - 
     r^4 \ell (\ell + 2) \right] R_{\bf III} &= 0 &
    \qquad\ &
     R_{\bf III} &= \alpha \df{J_{\ell+1}(\omega r)}{\omega r} + 
      \beta \df{N_{\ell+1}(\omega r)}{\omega r} \ , \cr
  }}}
 where $\alpha$ and $\beta$ are constants to be determined in the
matching and 
  \eqn\EtaDef{
   \eta = -\tf{1}{4} \sum_{i=1,5,K} {r_1 r_5 r_K \omega \over r_i^2}
        \equiv -{\omega \over 4 \pi T_L}
  }
 is the charge parameter of the Coulomb functions.  The infalling
solution $R_{\bf I}$ can be matched directly onto $R_{\bf III}$
without the aid of an intermediate region ${\bf II}$, with the result
  \eqn\EtaMatch{
   \alpha = {\ell! \over C_{\ell/2}(\eta)} 
    {2^{2\ell+1} \pi^\ell \over (A_{\rm h} \omega^3)^{\ell/2}} \ , 
   \quad \beta = 0 \ .
  }
 The quantity 
  \eqn\CDef{
   C_{\ell/2}(\eta) = {2^{\ell/2} e^{-\pi \eta / 2} 
    \Big| \Gamma\left( {\ell \over 2} + 1 + i \eta \right) \Big| \over
     \Gamma(\ell + 2)}
  }
 enters into the series expansion of Coulomb functions.  

A more accurate matching can be obtained with $\beta \neq 0$, but the
level of accuracy embodied in \EtaMatch\ is sufficient for the flux
ratio method \refs{\mast,\gkgrey,\cgkt}.  In this method, the
absorption probability is computed as the ratio of the infalling flux
at the horizon to the flux in the incoming wave at infinity.  The
result is
  \eqn\ProbAbs{
   1-|S_\ell|^2 = {1 \over \pi} {A_{\rm h} \omega^3 \over |\alpha|^2} 
    = 4 \pi \left( A_{\rm h} \omega^3 \over 
        16 \pi^2 \right)^{\ell+1} 
       {C_{\ell/2}^2(\eta) \over \ell!^2} \ .
  }
 Now the formula \SigmaAbsFF\ comes into play to give the final
result:
  \eqn\SigmaGR{\eqalign{
   \sigma_{\rm abs}^\ell &= A_{\rm h} (\ell + 1)^2 
    \left( A_{\rm h} \omega^3 \over 16 \pi^2 \right)^\ell
    {C_{\ell/2}^2(\eta) \over \ell!^2}  \cr
   &= {A_{\rm h} \over \ell!^4}
    \left( {A_{\rm h} \omega^3 \over 8 \pi^2} \right)^\ell
    e^{\omega \over 4 T_L} 
    {\textstyle \left| \Gamma\left( {\ell \over 2} + 1 - 
     i {\omega \over 4 \pi T_L} \right) \right|^2}
  }}
 The right hand side of \SigmaGR\ depends on $r_1$, $r_5$, and $r_K$
only through $A_{\rm h}$ and $T_L$.  Both quantities are symmetric in
the three radii, and in fact admit U-duality invariant generalizations
\finn.  The near-extremal generalization of \SigmaGR,
  \eqn\SigmaNonEx{
   \sigma_{\rm abs}^\ell = A_{\rm h} 
    {(\omega r_0 / 2 )^{2\ell} \over \ell!^4}
    \left| {\Gamma\left( 1 + {\ell \over 2} - 
                {i \omega \over 4 \pi T_L} \right)
            \Gamma\left( 1 + {\ell \over 2} - 
                {i \omega \over 4 \pi T_R} \right)
     \over \Gamma\left( 1 - i {\omega \over 2 \pi T_H} \right)}
    \right|^2 \ ,
  }
 also treats the three charges on an equal footing.  The temperatures
$T_L$ and $T_R$ are given by \finn
  \eqn\TLR{
   \beta_{L,R} = 2 \pi r_0 \left( \prod_i \cosh \sigma_i \mp
    \prod_i \sinh \sigma_i \right) 
  }
 in the general non-extremal case.

\newsec{The effective string analysis}
\seclab\Dbrane

Despite recent progress in formulating superspace actions for branes
\refs{\cederOne,\cederTwo,\aps} and in generalizing the DBI action to
nonabelian gauge theory (see \atWhich\ and references therein), a
first-principles derivation of a complete action for the effective
string, including all couplings to fields in the bulk of spacetime,
has yet to be achieved.  The goal of this section is to write down a
reasonable form for the part of the action responsible for coupling
the effective string to higher partial waves of an ordinary scalar and
see how the cross-sections it predicts compare with the semi-classical
result \SigmaGR.

Consider the off-diagonal graviton $h_{ij}$ with $i$ and $j$ parallel
to the D5-brane but perpendicular to the D1-brane.  The lowest-order
interaction of this field with excitations on the effective string can
be read off from the DBI action \dmOne: in static gauge where $t =
\tau$ and $x^5 = \sigma$,
  \eqn\VintSure{
   V_{\rm int} = -t_\eff \int_0^{L_\eff} d\sigma \,
    2 h_{ij}(\tau,\sigma,\vec{x}\!=\!0) 
     \partial_+ X^i \partial_- X^j \ .
  }
 The convention in \VintSure\ and elsewhere is to sum over all $i \neq
j$.  The fields $h_{ii}$ couple somewhat differently; an exploration
of those couplings and their physical consequences was initiated in
\refs{\cgkt}.
 
In \refs{\juan}, an analysis of the entropy and temperature of
near-extremal 5-branes led to an effective string with $c_\eff = 6$
and $T_\eff = 1/(2 \pi r_5^2)$.  An extension of the methods used in
\juan\ to the case $r_1 \sim r_5$ leads to \gforth 
  \eqn\CapTeff{
   T_\eff = {1 \over 2 \pi (r_1^2 + r_5^2)} \ .
  }
 The natural assumption is that $t_\eff$, by definition the tension
that appears in front of the DBI action, is precisely $T_\eff$.
Strangely enough, all previous scattering calculations except the
fixed scalar computation of \kkTwo\ (whose implications regarding the
effective string tension are unclear to me) either do not depend on
$t_\eff$ or require $r_1 = r_5$.  Thus, purely from the point of view
of scattering computations, $t_\eff$ seems ambiguous by a factor of
the form $f(r_1/r_5)$ where $f(1) = 1$.  One of the motivations for
studying higher partial waves is to resolve this ambiguity.  The
result I will obtain is 
  \eqn\tandT{
   t_\eff = {1 \over 2 \pi r_1 r_5} \ .
  }

Because of the evaluation of $h_{ij}$ at $\vec{x} = 0$, \VintSure\ is
a coupling to the $s$-wave of $h_{ij}$ only.  How might it be
generalized to include the dominant couplings to arbitrary partial
waves?  To begin with, a coupling to the $\ell^{\rm th}$ partial wave
should include $\ell$ derivatives of $h_{ij}$ since the wave-function
vanishes like $|\vec{x}|^\ell$.  It is the fermions on the effective
string which carry the angular momentum \refs{\brekEx,\brekNonEx}: the
left-moving and right-moving fermions transform in a fundamental of
$SU(2)_L$ and $SU(2)_R$ respectively, where the $SO(4)$ of rotations
in the four noncompact spatial dimensions is written as $SO(4) =
SU(2)_L \times SU(2)_R$.  Purely on group theory grounds, one thus
expects the $\ell^{\rm th}$ partial wave to couple to $\ell$
left-moving and $\ell$ right-moving fermions.  The order of the
absorption process in the string coupling can be read off from
\SigmaGR\ as $g^{\ell + 1}$ where $g$ is the closed string coupling.
Exactly two more open string vertex operators should be included in
the interaction to make the disk diagram come out with this power of
$g$.  The natural candidate for the interaction is
  \eqn\VintGuess{\eqalign{
   V_{\rm int} &= -t_\eff \int_0^{L_\eff} d\sigma \,
    2 h_{ij}(\tau,\sigma,x^m\!=\!\bar\Psi \gamma^m \Psi) 
     \partial_+ X^i \partial_- X^j  \cr
    &= -t_\eff \int_0^{L_\eff} d\sigma \,
    2 \sum_{\ell=0}^\infty {1 \over \ell!} 
     \left( \prod_{k=1}^\ell \bar\Psi \gamma^{m_k} \Psi \right)
     \partial_{m_1} \cdots \partial_{m_\ell} 
     h_{ij}(\tau,\sigma,x^m\!=\!0) \,
     \partial_+ X^i \partial_- X^j \ .
  }}
 Extra derivatives on the fermion fields are possible {\it a priori},
but power counting in $\omega$ for $\omega/T_L \ll 1$ shows that they
must be absent if \SigmaGR\ is to be reproduced.  The same general
form of the coupling was deduced independently in \ja\ through a
greybody factor analysis.

The outstanding fallacy of \VintGuess\ is that the sum terminates at
$\ell = 4$ because there are only four types of left-moving fermions
and the same number of right-moving fermions.  The situation is even
worse when one factors in the restrictions from $SO(4)$ group theory.
As we shall see after \VintGuessTwo, only the $\ell = 0$ and $\ell =
1$ partial waves can be absorbed.  

In \VintGuess, the $\gamma^m$ are gamma matrices of $SO(4,1)$:
  \eqn\GammaMs{
   \gamma^0 = \pmatrix{ -i & 0  \cr 
                         0 & i } \qquad
   \gamma^m = \pmatrix{ 0 & \tau^m_{\alpha\dot\beta}  \cr
                        \tau^{m\dot\alpha\beta} & 0 }
  }
 where 
  \eqn\TauMs{
   \tau^m_{\alpha\dot\alpha} = 
    \left( {\bf 1},i \sigma_1,i \sigma_2,i \sigma_3 \right)  \qquad
   \tau^{m\dot\alpha\alpha} = \epsilon^{\dot\alpha\dot\beta}
    \epsilon^{\alpha\beta} \tau^m_{\beta\dot\beta} =
    \left( {\bf 1},-i \sigma_1,-i \sigma_2,-i \sigma_3 \right) \ ,
  }
 $\sigma_i$ being the usual Pauli matrices.  I follow northwest
contraction conventions for raising and lowering spinor indices, and I
set $\epsilon_{01} = \epsilon^{01} = \epsilon_{\dot0\dot1} =
\epsilon^{\dot0\dot1} = 1$.  The four-component spinor $\Psi$
decomposes into $SU(2)_L$ and $SU(2)_R$ fundamentals,
which are left-movers and right-movers on the effective string,
respectively.  These complex fermions decompose further into real
components of the 10-dimensional Majorana-Weyl spinor that one would
expect to emerge most simply from a full string theory analysis:
  \eqn\PsiDs{\vcenter{\openup1\jot
   \halign{\strut\span\TC\cr
   \Psi = \pmatrix{ \Psi_{+\alpha} \cr \bar\Psi_-^{\dot\alpha} }  \cr
    \Psi_\pm^1 = {\psi_\pm^1 + i \psi_\pm^2 \over \sqrt{2}} \qquad
    \Psi_\pm^2 = {\psi_\pm^3 + i \psi_\pm^4 \over \sqrt{2}} \ . \cr
  }}}
 Note that complex conjugation raises or lowers a spinor index, rather
than dotting or undotting it as in the case of $SO(3,1)$.

For $\ell \ge 2$ the $\ell^{\rm th}$ term in the interaction
\VintGuess\ makes subleading contributions to the absorption of lower
partial waves because the expression $\partial_{m_1} \cdots
\partial_{m_\ell} h_{ij}$ does not pick out a pure $\ell^{\rm th}$
partial wave from a plane wave.  The cure for this is to symmetrize
$SU(2)_L$ and $SU(2)_R$ spinor indices:
  \eqn\VintGuessTwo{\eqalign{
   V_{\rm int} &= -2 t_\eff \int_0^{L_\eff} d \sigma \,
    \sum_{\ell=0}^4 {i^\ell \over \ell!} 
    \prod_{k=0}^\ell \left( \Psi_+^{\alpha_k} \Psi_{-\dot\beta_k} + 
     \bar\Psi_+^{\alpha_k} \bar\Psi_{-\dot\beta_k} \right)
    \left( \tau^{m_1 (\dot\beta_1}_{(\alpha_1} \cdots
     \tau^{|m_\ell|\dot\beta_\ell)}_{\alpha_\ell)} \right)  \cr
   &\qquad\quad \cdot \partial_{m_1} \cdots \partial_{m_\ell} h_{ij}
    \partial_+ X^i \partial_- X^j + \ldots  
  }}
 Terms have been omitted in \VintGuessTwo\ which make subleading
contributions.  The product of fermion fields is antisymmetric in
$\alpha_1 \ldots \alpha_\ell$ and in $\dot\beta_1 \ldots
\dot\beta_\ell$.  Hence all terms in \VintGuessTwo\ vanish except
$\ell = 0$ and $\ell = 1$.  The conclusion is that only the first two
partial waves can be absorbed.  This limitation is not merely a
failing of the specific form \VintGuess; it is intrinsic to the
approach of coupling partial wave to a product of fermion fields
evaluated at a single point on the effective string without
derivatives.  To reiterate, the addition of derivatives introduces
extra powers of the energy in the final cross-section which would
cause disagreement with \SigmaGR.  Let us proceed with the analysis of
$\ell = 1$ and consider possible extensions to $\ell \ge 2$ later.

There are many steps involved in passing from the interaction
\VintGuessTwo\ to the cross-section for $\ell = 1$.  To avoid losing
factors it pays to be as explicit as possible.  Let us begin with mode
expansions of the fields appearing in \VintGuessTwo.  The forms of mode
expansions are dictated by the kinetic terms in the action.  In the
present case, the kinetic terms are
  \eqn\SKinetic{\eqalign{
   S_{\rm bulk} &= {1 \over 2 \kappa_6^2} \int d^6 x \,
    \tf{1}{4} (\partial_\mu h_{ij}) (\partial^\mu h^{ij}) + \ldots \cr
   S_{\rm string} &= -2 t_\eff \int d^2 \sigma \, 
    \left[
     \partial_+ X_i \partial_- X^i + 
     \psi_+^\Delta i \partial_- \psi_+^\Delta + 
     \psi_-^{\dot\Delta} i \partial_+ \psi_-^{\dot\Delta} + 
    \ldots \right] \ ,
  }}
 resulting in the mode expansions
  \eqn\ModeExpansions{\eqalign{
   \psi_+^\Delta(\tau + \sigma) &= 
    \sum_{k^5 \in {2 \pi \over L_\eff} ({\bf Z}^- - 1/2)}
     {1 \over \sqrt{2 L_\eff t_\eff}}
     \left( b_k^\Delta e^{i k \cdot \sigma} + \hc \right) \cr
   \psi_-^{\dot\Delta}(\tau - \sigma) &= 
    \sum_{k^5 \in {2 \pi \over L_\eff} ({\bf Z}^+ + 1/2)}
     {1 \over \sqrt{2 L_\eff t_\eff}}
     \left( b_k^{\dot\Delta} e^{i k \cdot \sigma} + \hc \right) \cr
   X^i(\sigma,\tau) &= 
    \sum_{k^5 \in {2 \pi \over L_\eff} {\bf Z}}
     {1 \over \sqrt{2 L_\eff t_\eff k^0}}
     \left( a_k^i e^{i k \cdot \sigma} + \hc \right) \cr
   h^{ij}(x^\mu) &= 
    \sum_{k^5,\vec{k}}
     \sqrt{2 \kappa_6^2 \over 2 V L_5 k^0}
     \left( g_k^{ij} e^{i k \cdot x} + \hc \right) \ .
  }}
 The sum over $k^5,\vec{k}$ has $k^5 \in {2 \pi \over L_5} {\bf Z}$
and $k^m \in {2 \pi \over \root{4}\of{V}} {\bf Z}$ for $m = 1,2,3,4$.
$V$ is the volume of a large box in which we imagine enclosing the
four uncompactified spatial dimensions.  The indices $\Delta$ and
$\dot\Delta$ run from $1$ to $4$, and since $\psi^\Delta_+$ and
$\psi^{\dot\Delta}_-$ are real, conjugation does not change the
position of the indices.  Typographical convenience will dictate the
position of $\Delta$ and $\dot\Delta$.  

It is important that $h^{ij}$ is moded differently in the $x^5$
direction from the effective string excitations: the minimal quantum
of Kaluza-Klein charge for an excitation on the effective string is
$1/(n_1 n_5)$ of the minimal quantum for a particle in the bulk
\refs{\dmI,\ms}.  In \ModeExpansions\ we have not been careful about
zero modes because it is the oscillator states which are important for
the absorption processes.  The factors in \ModeExpansions\ were chosen
to make the commutation relations simple:
  \eqn\ModeCommutators{\eqalign{
   \{ b_k^\Delta,{b_q^\Gamma}\+ \} &= 
      \delta_{k^5-q^5} \delta^{\Delta\Gamma} \cr
   \{ b_k^{\dot\Delta},{b_q^{\dot\Gamma}}\+ \} &= 
      \delta_{k^5-q^5} \delta^{\dot\Delta\dot\Gamma} \cr
   [a_k^i,{a_q^j}\+] &= 
      \delta_{k^5-q^5} \delta^{ij} \cr
   [g_k^{ij},{g_q^{fh}}\+] &= 
      \delta_{\vec{k}-\vec{q}} \delta^{if} \delta^{jh} \quad
      \hbox{if $i<j$ and $f<h$} \ .
  }}

The goal now is to compute the amplitude 
  $\langle \tilde{f} | V_{\rm int} | \tilde{i} \rangle$
 for an absorption process where a scalar in the $\ell = 1$
partial wave turns into two bosons and two fermions on the
effective string.  The tildes on $| \tilde{i} \rangle$ and $|
\tilde{f} \rangle$ are meant to indicate that these state vectors are
not the real initial and final states: they include only the particles
that participate in the interaction and not the whole thermal sea of
left-movers that give the effective string its Kaluza-Klein charge.
Restoring the thermal sea is an easy exercise which will be postponed
until \EnhancedGolden.

Two other slight simplifications will be made to ease the notational
burden.  First, indices can be dropped on all the $X^i$ fields, but
then one must include an extra factor of $2$ in the rate, as shown in
\ONZTEnhancedGolden.  The $2$ accounts for the fact that $h_{ij}$ can
turn into a left-moving $X^i$ and a right-moving $X^j$ or a
left-moving $X^j$ and a right-moving $X^i$.  The second simplification
is to consider only
  \eqn\TildeIF{\eqalign{
   |\tilde{i}\rangle &= g_k\+ |0\rangle \cr
   |\tilde{f}\rangle &= a_{p_b}\+ a_{q_b}\+ 
    {b_{p_f}^{\Delta}}\+ {b_{q_f}^{\dot\Delta}}\+ |0\rangle \ ,
  }}
 which is to say we put all the particles on the effective string into
the final state and none into the initial state.  A simple way to
account for all the crossed processes which also contribute to
absorption will be discussed after equation \ILeft.  In \TildeIF\
and below, $k$ refers to the momentum of the bulk scalar, $p$ refers
to the momentum of a left-mover on the effective string, and $q$
refers to the momentum of a right-mover.

The desired matrix element can now be read off from \VintGuessTwo\
and \ModeCommutators\ as 
  \eqn\OneNZTerm{
   \langle\tilde{f}| V_{\rm int} |\tilde{i}\rangle =
    C_\Delta^{\dot\Delta}
     {k_1 \over 2 L_\eff t_\eff} \kappa_5
     \sqrt{p_b^0 q_b^0 \over V k^0}
     \delta_{k^5 - p_b^5 - p_f^5 - q_b^5 - q_f^5} \ ,
  }
 where $C_\Delta^{\dot\Delta}$ is the $4 \times 4$ matrix
$\diag\{1,-1,1,-1\}$.  To extract the rate is is necessary to use a
generalization of Fermi's Golden Rule that includes Bose enhancement
factors and Fermi suppression factors:
  \eqn\ONZTEnhancedGolden{\eqalign{
   \Gamma &= 2 \sum_{|\tilde{f}\rangle} 
    (\rho_L^X(p_b^0) + 1) (\rho_R^X(q_b^0) + 1) 
    (1 - \rho_L^\psi(p_f^0)) (1 - \rho_L^\psi(q_f^0))  \cr
    &\qquad \cdot \left| \langle\tilde{f}| V_{\rm int} 
      |\tilde{i}\rangle \right|^2 
      2 \pi \delta\left( k^0 - p_b^0 - p_f^0 - 
         q_b^0 - q_f^0 \right)  \cr
    &= 2 \sum_{\Delta,\dot\Delta} 
      | C_\Delta^{\dot\Delta} |^2  
     \cdot \sum_{\rm modes}
       (\rho_L^X(p_b^0) + 1) (\rho_R^X(q_b^0) + 1) 
       (1 - \rho_L^\psi(p_f^0)) (1 - \rho_L^\psi(q_f^0))  \cr
    &\qquad \cdot 
       {k_1^2 \kappa_5^2 \over V k^0
        (2 L_\eff t_\eff)^2} p_b^0 q_b^0 
        \delta_{k^5 - p_b^5 - p_f^5 - q_b^5 - q_f^5}
        2 \pi \delta\left( k^0 - p_b^0 - p_f^0 - 
         q_b^0 - q_f^0 \right)
  }}
 where
  \eqn\Rhos{
   \rho_L^X(p_b^0) = {1 \over e^{p_b^0 / T_L} - 1} \ , \qquad
   \rho_L^\psi(p_i^0) = {1 \over e^{p_i^0 / T_L} + 1}
  }
 and similarly for the right-moving thermal occupation factors.
Again, the explicit factor of $2$ in the first line of
\ONZTEnhancedGolden\ is present to account for the two distinct 
choices, $i+$~$j-$ or $i-$~$j+$, for polarizing the bosonic fields.  

The vanishing of all cross-sections beyond $\ell = 1$ is an egregious
failing of the most naive effective string model.  The simplest fix
would be to allow an incoming scalar to couple to a product of fermion
fields evaluated at a single point on the spatial $S^1$ which the
effective string wraps, but not necessarily at a single point in the
effective string coordinates $\sigma$.  In terms of the (4,4) SCFT
from which the effective string emerges as a particular twisted
sector, this more general coupling seems very natural because it still
involves only a local operator constructed from a product of the $4
n_1 n_5$ species of fermions in the SCFT.

The present treatment extends easily to cover this more general
interaction.  Let $D_k$ and $\dot{D}_k$ be $4 n_1 n_5$-valued indices
for the left- and right-moving fermion fields, respectively.  Consider
the final state
  \eqn\MoreTildeF{
   |\tilde{f}\rangle = a_{p_b}\+ a_{q_b}\+ 
    \left( {b_{p_1}^{D_1}}\+ {b_{q_1}^{\dot{D}_1}}\+ \cdots
           {b_{p_\ell}^{D_\ell}}\+ {b_{q_\ell}^{\dot{D}_\ell}}\+ 
    \right) |0\rangle \ .
  }
 The matrix element is now of the form 
  \eqn\NonZeroTerms{
   \langle\tilde{f}| V_{\rm int} |\tilde{i}\rangle =
    C_{D_1 \ldots D_\ell}^{\dot{D}_1 \ldots \dot{D}_\ell} 
     {k_1^\ell \over (2 L_\eff t_\eff)^\ell} \kappa_5
     \sqrt{p_b^0 q_b^0 \over V k^0}
     \delta_{k^5 - p_b^5 - \Sigma p_i^5 - q_b^5 - \Sigma q_i^5} \ .
  }
 The coefficient tensor $C_{D_1 \ldots D_\ell}^{\dot{D}_1 \ldots
\dot{D}_\ell}$ is antisymmetric in $D_1 \ldots D_\ell$ and in
$\dot{D}_1 \ldots \dot{D}_\ell$.  It encodes the $SO(4)$ group theory
factors isolating the $\ell^{\rm th}$ partial wave as well as
restrictions on the possible final states arising from D1-brane and
D5-brane Chan-Paton factors.  The rate is
  \eqn\EnhancedGolden{\eqalign{
   \Gamma &= 2 \sum_{|\tilde{f}\rangle} 
    (\rho_L^X(p_b^0) + 1) (\rho_R^X(q_b^0) + 1) 
    \prod_{i=1}^\ell \left[ (1 - \rho_L^\psi(p_i^0)) 
           (1 - \rho_L^\psi(q_i^0)) \right]  \cr
    &\qquad \cdot \left| \langle\tilde{f}| V_{\rm int} 
      |\tilde{i}\rangle \right|^2 
      2 \pi \delta\left( k^0 - p_b^0 - {\textstyle\sum} p_i^0 - 
         q_b^0 - {\textstyle\sum} q_i^0 \right)  \cr
    &= {2 \over \ell!^2} \sum_{D_k,\dot{D}_k} 
      | C_{D_1 \ldots D_\ell}^{\dot{D}_1 \ldots 
         \dot{D}_\ell} |^2  
     \cdot \sum_{\rm modes}
       (\rho_L^X(p_b^0) + 1) (\rho_R^X(q_b^0) + 1) 
       \prod_{i=1}^\ell \left[ (1 - \rho_L^\psi(p_i^0)) 
              (1 - \rho_L^\psi(q_i^0)) \right]  \cr
    &\qquad \cdot 
       {k_1^{2 \ell} \kappa_5^2 \over V k^0
        (2 L_\eff t_\eff)^{2\ell}} p_b^0 q_b^0 
        \delta_{k^5 - p_b^5 - \Sigma p_i^5 - q_b^5 - \Sigma q_i^5}
        2 \pi \delta\left( k^0 - p_b^0 - {\textstyle\sum} p_i^0 - 
         q_b^0 - {\textstyle\sum} q_i^0 \right) \ .
  }} 
 The $1/\ell!^2$ in the last expression arises because the sums over
$D_k$, $\dot{D}_k$, $p_k$, and $q_k$ are unrestricted, and there are
$\ell!^2$ different permutations of a given set of values for these
quantities which yield the same final state $|\tilde{f}\rangle$.

When the energy of the incoming scalar is much greater than the gap, a
continuum approximation can be made in \EnhancedGolden:
  \eqn\ContApprox{
   \sum_p \to \int dp \, {L_\eff \over 2 \pi} \ , \qquad
   \delta_p \to {2 \pi \over L_\eff} \delta(p) \ .
  }
 For the sake of simplicity, only massless particle absorption will be
considered.  In that case the flux associated with the state
$g_k\+|0\rangle$ is ${\cal F} = 1/V$.  The absorption cross-section is
  \eqn\SigmaPLR{\eqalign{
   \sigma_\abs &= 
     V \Gamma(scalar \to b_L + b_R + \ell f_L + \ell f_R) + 
     \hbox{crossed processes} \cr
    &= {\sum_{D_k,\dot{D}_k} 
      | C_{D_1 \ldots D_\ell}^{\dot{D}_1 \ldots 
         \dot{D}_\ell} |^2 \over \ell!^2 4^\ell} 
      {\kappa_5^2 L_\eff \over (2 \pi t_\eff)^{2 \ell}}
       \omega^{2 \ell - 1} I_L I_R
  }}
 where
  \eqn\ILeft{
   I_L = \int_{-\infty}^\infty d p_b^0 \prod_{i=1}^\ell d p_i^0 \,
    \delta\left( {\omega \over 2} - p_b^0 - 
     \sum_{i=1}^\ell p_i^0 \right) 
    {p_b^0 \over 1 - e^{-p_b^0 / T_L}}
    \prod_{i=1}^\ell {1 \over 1 + e^{-p_i^0 / T_L}}
  }
 and similarly for $I_R$.  
  $\Gamma(scalar \to b_L + b_R + \ell f_L + \ell f_R)$ 
 is what was computed in \EnhancedGolden.  Arbitrary crossings of the
basic process
  $scalar \to b_L + b_R + \ell f_L + \ell f_R$ 
 and their time-reversals also contribute to the net absorption rate
from which $\sigma_\abs$ is computed.  However, the simple trick of
extending the integrals in \ILeft\ over the entire real line can be
used to keep track of all of them.  A demonstration of this with
careful attention paid to symmetry factors can be found in
\refs{\cgkt} for the special case where only two left-moving bosons
and two right-moving bosons are involved.  Note that for $\ell = 0$
the dependence on $t_\eff$ disappears in \SigmaPLR, as was noted
previously in \refs{\cgkt}\fixit{And \dm?}.

The integral \ILeft\ is a convolution of $\ell + 1$ simple functions
and so can be done most directly by transforming to Fourier space,
where convolutions become products.  Three integrals which are useful
for doing the Fourier transforms are
  \eqn\SechIntegrals{\eqalign{
   \int_{-\infty}^\infty dp \, e^{i x p} {p \over 2 \sinh {p \over 2 T_L}}
    &= (\pi T_L)^2 {\rm sech}^2 \, (\pi T_L x) \cr
   \int_{-\infty}^\infty dp \, e^{i x p} {1 \over 2 \cosh {p \over 2 T_L}}
    &= (\pi T_L) {\rm sech} \, (\pi T_L x) \cr
   \int_{-\infty}^\infty dx \, e^{-i x p} (\pi T_L)^{\ell+2} 
    \sech^{\ell+2} (\pi T_L x) &=
    (2 \pi T_L)^{\ell+1} 
     {\Big| \Gamma\left( {\ell \over 2} + 1 - i {p \over 2 \pi T_L}
      \right) \Big|^2 \over (\ell+1)!} \ .
  }}
 The first two integrals are Fourier inversions of the third in the
special cases $\ell = 0$ and $-1$.  Now the computation is
straightforward: 
  \eqn\ILeftComp{\eqalign{
   I_L &= e^{\omega \over 4 T_L} 
     \int_{-\infty}^\infty d p_b^0 \prod_{i=1}^\ell d p_i^0 \,
     \delta\left( {\omega \over 2} - p_b^0 - 
      \sum_{i=1}^\ell p_i^0 \right) 
     {p_b^0 \over 2 \sinh {p_b^0 \over 2 T_L}}
     \prod_{i=1}^\ell {1 \over 2 \cosh {p_i^0 \over 2 T_L}} \cr
    &= {e^{\omega \over 4 T_L} \over 2 \pi}
     \int_{-\infty}^\infty dx e^{-i x \omega / 2} 
      (\pi T_L)^{\ell+2} \sech^{\ell+2}(\pi T_L x) \cr
    &= {(\ell+1)! \over \pi} (\pi T_L)^{\ell+1} C_{\ell/2}^2(\eta) \ .
  }}
 The last step uses \EtaDef\ and \CDef.  $I_R$ can be computed
similarly, but since $T_R = 0$ by assumption, the result is much
simpler:
  \eqn\IRight{
   I_R = {(\omega / 2)^{\ell+1} \over (\ell+1)!} \ .
  }

Now the absorption cross-section can be given in closed form:
  \eqn\SigmaD{\eqalign{
    \sigma_{\rm abs}^\ell &= {\sum_{D_k,\dot{D}_k} 
      | C_{D_1 \ldots D_\ell}^{\dot{D}_1 \ldots 
         \dot{D}_\ell} |^2 \over \ell!^2 4^\ell} 
      {\kappa_5^2 L_\eff \over (2 \pi t_\eff)^{2 \ell}}
      {\omega^{3 \ell} \over \pi}
      \left( {\pi T_L \over 2} \right)^{\ell+1} C_{\ell/2}^2(\eta) \cr
     &= A_{\rm h} \sum_{D_k,\dot{D}_k} 
      | C_{D_1 \ldots D_\ell}^{\dot{D}_1 \ldots 
         \dot{D}_\ell} |^2 
      \left( {A_{\rm h} \omega^3 \over 16 \pi^2} \right)^\ell
       {C_{\ell/2}^2(\eta) \over \ell!^2} \ .
  }}
 In the second equality two key relations have been used: 
  \eqn\TLt{
   T_L = {r_K \over \pi r_1 r_5}  \qquad
   {\kappa_5^2 L_\eff T_L \over 2} = A_{\rm h} \ .
  }
 Both are valid when $r_K \ll r_1,r_5$.  The first can be derived by
setting the effective string entropy equal to the Bekenstein-Hawking
entropy.  The second is a limiting case of \TLR.  In addition, the
tension has at last been fixed:
  \eqn\FixedTension{
   t_\eff = {1 \over 2 \pi r_1 r_5} \ .
  }
 Because the cross-section \SigmaGR\ depends on $n_1$ and $n_5$ only
through the product $n_1 n_5$ in the dilute gas regime, and because
the same is true of the quantities $L_\eff$, $T_L$, and $A_{\rm h}$
when $r_0 = 0$ and $r_K \ll r_1,r_5$, the choice $t_\eff \sim
1/\sqrt{n_1 n_5}$ seems inevitable.

Precise agreement between General Relativity and the effective string
now depends only on the relation
  \eqn\CNorm{
   \sum_{D_k,\dot{D}_k} 
      | C_{D_1 \ldots D_\ell}^{\dot{D}_1 \ldots 
         \dot{D}_\ell} |^2 = (\ell + 1)^2 \ .
  }
 Agreement in the case $\ell = 0$ is trivial.  For $\ell = 1$, the
original treatment in terms of free fermions on the effective string
is adequate: one can easily trace through the computations and
verify that $D_1$, $\dot{D}_1$, and $C_{D_1}^{\dot{D}_1}$ can be
replaced in every equation by $\Delta$, $\dot\Delta$, and
$C_\Delta^{\dot\Delta}$.  Any numerical discrepancy could have been
fixed by introducing a multiplicative constant in the relation $x^m =
\bar\Psi \gamma^m \Psi$; however to see perfect agreement without such
artifice is pleasing and also rather suggestive of the form one
expects for a gauge-fixed kappa symmetric action.

The real test is $\ell \ge 2$.  Here it seems essential to depart from
the simplistic effective string picture and return to a more
fundamental description of the D1-D5 bound state in order to compute
the coefficients $C_{D_1 \ldots D_\ell}^{\dot{D}_1 \ldots
\dot{D}_\ell}$.

Finally, it is worth noting that if agreement can be established for
extremal absorption, agreement for the near-extremal case follows
automatically.  In the effective string computation for near-extremal
absorption one must subtract off the stimulated emission contribution
as described in \cgkt\ in order to respect detailed balance and time
reversal invariance.  Modulo this subtraction, the result can be read
off from \SigmaPLR\ and \ILeftComp:
  \eqn\SigmaDNonEx{\eqalign{
   \sigma_{\rm abs}^\ell &= {\sum_{D_k,\dot{D}_k} 
      | C_{D_1 \ldots D_\ell}^{\dot{D}_1 \ldots 
         \dot{D}_\ell} |^2 \over \ell!^2 (\ell+1)!^2}
    {\kappa_5^2 L_\eff T_L T_R \over T_H} 
    \left( {\omega \sqrt{T_L T_R} \over 2 t_\eff} \right)^{2 \ell}  \cr
   &\qquad \cdot \left| {\Gamma\left( 1 + {\ell \over 2} - 
                {i \omega \over 4 \pi T_L} \right)
            \Gamma\left( 1 + {\ell \over 2} - 
                {i \omega \over 4 \pi T_R} \right)
     \over \Gamma\left( 1 - i {\omega \over 2 \pi T_H} \right)}
    \right|^2  \cr
   &= {\sum_{D_k,\dot{D}_k} 
      | C_{D_1 \ldots D_\ell}^{\dot{D}_1 \ldots 
         \dot{D}_\ell} |^2 \over \ell!^2 (\ell+1)!^2}
      A_{\rm h} \left( {\omega r_0 \over 2} \right)^{2 \ell}
    \left| {\Gamma\left( 1 + {\ell \over 2} - 
                {i \omega \over 4 \pi T_L} \right)
            \Gamma\left( 1 + {\ell \over 2} - 
                {i \omega \over 4 \pi T_R} \right)
     \over \Gamma\left( 1 - i {\omega \over 2 \pi T_H} \right)}
    \right|^2 \ .
  }}
 The second line relies on a modified version of \TLt\ applicable to
the near-extremal case with $r_0,r_K \ll r_1,r_5$:
  \eqn\TLtNonEx{
   \sqrt{T_L T_R} = {r_0 \over 2 \pi r_1 r_5} \qquad
   {\kappa_5^2 L_\eff T_L T_R \over T_H} = A_{\rm h} \ .
  }
 The same tension \FixedTension\ and the same relation \CNorm\
establish agreement between \SigmaDNonEx\ and \SigmaNonEx.

It has been suggested \refs{\finn} that the effective string picture
can be used to describe in a U-duality invariant fashion black holes
with arbitrary charges, possibly even far from extremality.  The
expression on the first line of \SigmaDNonEx\ depends only only $T_L$,
$T_R$, $T_H$, and $L_\eff$---all quantities that have meaning to the
effective string considered in the abstract, independent of the
microscopic D1-D5-brane model.  It cries out to be reconciled
with \SigmaNonEx\ for arbitrary values of $r_0$, $Q_1$, $Q_5$, and
$Q_K$.  But the treatment of absorption given in this section relies
on the dilute gas approximation and thus is not general enough to be
matched in any meaningful way to General Relativity when $Q_K
\ll\!\!\!\!\!/\,\,\, Q_1,Q_5$.

\global\advance\secno by1\message{(\the\secno. Here's an awful hack
to get footnotes on section headers which works only with sequential 
numbering of equations)}
\global\subsecno=0\bigbreak\noindent{\bf\the\secno. 
Limitations on partial wave absorption\foot{The ideas
in this section originate largely in discussions with C.~Callan,
L.~Thorlacius, and J.~Maldacena.}}
\writetoca{{\secsym} {Limitations on partial wave 
absorption}}\par\nobreak\medskip\nobreak
\seclab\llimits

Consider the more general couplings described in the paragraph
preceding \MoreTildeF: local on spatial $S^1$ wrapped by the effective
string but not on the effective string itself.  Although the details
of the $SO(4)$ group theory and Chan-Paton factors have yet to be
worked out fully in the context of the (4,4) SCFT description of the
D1-D5-brane bound state, it seems clear that a coupling of the
$\ell^{\rm th}$ partial wave to an operator built out of any
combination of the $4 n_1 n_5$ fermionic fields raises the maximum
value of $\ell$ which the effective string can absorb from $1$ to some
number on the order $n_1 n_5$.  One might suppose that by putting
derivatives on some of the fermion fields, the problem can be avoided
altogether.  But such derivatives raise the dimension of the operator
and hence suppress the cross-section by more powers of $\omega$ than
are present in the semi-classical result.  To sum up, the assumption
of locality prevents the effective string from coupling to partial
waves above a certain maximum $\ell$ with the strength required to
match General Relativity.  The situation does not seem as satisfactory
as for the D3-brane, where couplings to all partial waves exist with
appropriate dimensions to reproduce semi-classical cross-sections
\refs{\klebThree} (normalizations however are problematic
\refs{\gukt}).

There is a reason, however, why one might expect not to observe
agreement between the effective string and General Relativity at high
values of $\ell$.  If the black hole absorbs some very high partial
wave, it winds up with a large angular momentum, so the geometry
before and after is appreciably different.  Back reaction is not
included in the General Relativity calculations of section~\semicl.
In fact the only back reaction calculation I am aware of for the black
hole under consideration \kvk\ is restricted to $\ell = 0$.  But on
the grounds of cosmic censorship one would expect that absorption
processes which drive the black hole past extremality are forbidden
even semi-classically.  From an adaptation of the work of
\refs{\brekEx,\brekNonEx} one can read off the corresponding bound on
$\ell$ as $\ell \lsim \sqrt{n_K n_1 n_5}$.

We have two different bounds on $\ell$ indicating the maximum partial
wave that the effective string should be capable of absorbing:
  \eqn\ThreeBounds{\vcenter{\openup1\jot
    \halign{\strut\span\TL & \span\TR & \qquad\span\TT\cr
    \ell &\lsim \ell_{\rm max}^L \equiv n_1 n_5 & 
      from locality and statistics  \cr
    \ell &\lsim \ell_{\rm max}^C \equiv \sqrt{n_K n_1 n_5} & 
      from cosmic censorship.  \cr
  }}}
 Now I would like to inquire which is the more restrictive.  Using the
standard relations (see for example \gkOne)
  \eqn\NRels{
   n_1 n_5 = {4 \pi^3 r_1^2 r_5^2 \over \kappa_5^2 L_5} \qquad\quad
   n_K = {\pi L_5 r_K^2 \over \kappa_5^2}
  }
 and the formula \TLt\ for $T_L$, one can show that $\ell_{\rm
max}^C/\ell_{\rm max}^L = L_5 T_L / 2$.

The validity of any comparison between General Relativity and the
effective string model as treated in section~\Dbrane\ relies on being
in the dilute gas regime \cm\ and at low energies \juanLow:
  \eqn\StandardLims{
   r_K \ll r_1,r_5 \ll 1/\omega \ .
  }
 These inequalities still do not determine whether $\ell_{\rm max}^L$
is larger or smaller than $\ell_{\rm max}^C$.  But if it is agreed to
examine only fat black holes \ms, which is to say if one assumes
  \eqn\FatLim{
   L_5 \ll \sqrt{r_1 r_5} \ ,
  }
 then it is easy to obtain the inequality
  \eqn\DesiredLims{
   \ell_{\rm max}^L \gg \ell_{\rm max}^C
  }
 by combining the dilute gas inequality in \StandardLims\ with
\FatLim.  Now, \DesiredLims\ is a hopeful state of affairs for the
effective string model, because it indicates that making couplings
local only on $S^1$ in principle enables the effective string to
absorb all the partial waves for which reasonable comparisons can be
made with General Relativity.  It is perhaps not the ideal state of
affairs: one might have hoped that the bounds $\ell_{\rm max}^L$ and
$\ell_{\rm max}^C$ would coincide, indicating that the effective
string knew about cosmic censorship.  The results of \ja\ suggest that
a more careful treatment of these issues using techniques of conformal
field theory would result in a translation of cosmic censorship into
unitarity of the effective string description.  Such a treatment would
need to address the problem that if one moves deep into the black
string region of parameter space by making $L_5$ large, one can obtain
$\ell_{\rm max}^L \ll \ell_{\rm max}^C$.  The perturbative D-brane
region is in fact closer to the black string region than the fat black
hole region, so it would be surprising to find such a disaster for
comparisons in the black string region when agreement seems possible
for fat black holes.

\newsec{Conclusion}
\seclab\Conclusion

In this paper I have shown that the leading order coupling of the
effective string to an ordinary scalar correctly predicts the
cross-sections for the $\ell=1$ partial wave.  Due to the Grassmannian
character of the fermionic fields which carry the angular momentum, it
is impossible for the simplest effective string model (a single long
string with $c_\eff = 6$) to couple properly to $\ell>1$ partial waves
through an operator local on the string.  Generalizing the model to
include what one might intuitively regard as multi-strand interactions
of the effective string postpones this difficulty to $\ell \gsim n_1
n_5$, a higher bound on $\ell$ for fat black holes than the one
arising from cosmic censorship.  An analysis of the unique form which
such interactions must have in order to make a leading order
contribution to the absorption of the $\ell^{\rm th}$ partial wave
demonstrates that the correct energy dependence arises from the finite
temperature kinematics.  This demonstration, together with the general
proof of the Optical Theorem for absorption of scalars given in
section~\semicl, can be viewed as a full investigation of the
kinematics involved in higher partial waves.  What is left is to
calculate the coefficients $C_{D_1 \ldots D_\ell}^{\dot{D}_1 \ldots
\dot{D}_\ell}$ and thereby verify or falsify \CNorm, on which
agreement between General Relativity and the effective string model
relies.

The means to achieve a clear description of the dynamics and hopefully
a derivation of the $C_{D_1 \ldots D_\ell}^{\dot{D}_1 \ldots
\dot{D}_\ell}$ is a more precise treatment of the low-energy SCFT
dictating the dynamics of the D1-D5-brane bound state.  The effective
string might continue to be a useful picture, perhaps supplemented by
rules governing how different strands of the effective string
interact.  I hope to report on this approach in the future.

In a way the finding that the effective string tension needs to scale
as $1/\sqrt{Q_1 Q_5}$ is a more serious difficulty than the vanishing
of $\ell > 1$ cross sections.  Indeed, the necessity of choosing this
peculiar value for the tension appears already at $\ell = 1$, where
the naive effective string model in other ways seems completely
adequate.  A study of T-duality in \mathur\ led to the conclusion that
a tension scaling as $1/\sqrt{Q_1 Q_5}$ is more natural than $1/Q_1$
or $1/Q_5$, and it was further described how a simple modification in
the calculation of disk diagrams would lead to such a scaling.
However, in the absence of a first-principles derivation of the
tension, a $1/(Q_1+Q_5)$ scaling seems equally natural.  This scaling
is the one favored by entropy and temperature arguments.  Thus it
appears that a single energy scale does not fully characterize
effective strings in the way that $\alpha'$ does fundamental strings.

Although effective string models of black holes have recently enjoyed
a number of remarkable successes, a unifying picture has been slow in
emerging.  The General Relativity calculations for near-extremal black
holes, on which most of the evidence for effective strings is based,
are conceptually straightfoward.  The difficulty of studying bound
states of solitons has caused the link between fundamental string
theory and effective strings to remain imprecise in certain
respects---most importantly in the interaction between the effective
string and fields in the bulk of spacetime.

The puzzles presented by higher partial waves may push effective
string theory in the directions it needs to go in order to become a
fully viable model of near-extremal black holes.  Discrepant results
for the tension may be a clue to nature of effective string's
interactions with bulk fields.  Rescuing the $\ell > 1$ cross-sections
must surely lead to a consideration of the multi-strand interactions
which are as yet virgin territory in the theory of effective strings.
Already, the absorption of higher partial waves into five-dimensional
black holes is the cleanest dynamical test of the role of fermions on
the effective string.  Achieving a full understanding of these
processes would constitute a major advance, not only in establishing
the viability of effective strings in black hole physics, but also in
comprehending the D1-D5-brane bound state.


\bigbreak\bigskip\bigskip\centerline{{\bf Acknowledgements}}\nobreak

  I would like to thank I.~Klebanov, C.~Callan, F.~Larsen,
J.~Maldacena, S.~Mathur, W.~Taylor, L.~Thorlacius, and V.~Periwal for
useful discussions.  This work was supported in part by DOE grant
DE-FG02-91ER40671 and the NSF Presidential Young Investigator Award
PHY-9157482.  I also thank the Hertz Foundation for its support.

\listrefs
\bye